\documentclass[preprintnumbers, prd, showpacs, floatfix,
preprintnumbers, letterpaper, superscriptaddress]{revtex4}
\usepackage{eurosym}
\usepackage{amsfonts}
\usepackage{amsmath}
\usepackage{amssymb,epsf}
\usepackage{latexsym}
\usepackage{graphicx,epsfig}
\usepackage{amssymb}
\usepackage{subfigure}
\usepackage[colorlinks=true,citecolor=blue,linkcolor=blue,urlcolor=black]{hyperref}
\usepackage{epstopdf}
\usepackage{xcolor}
\usepackage{url}

\newcommand{\ba}{\begin{align}}
\newcommand{\ea}{\end{align}}
\newcommand{\bq}{\begin{equation}}
\newcommand{\eq}{\end{equation}}
\newcommand{\barr}{\begin{eqnarray}}
\newcommand{\earr}{\end{eqnarray}}

\begin{document}

\title{Alternative approach to the critical behavior and microscopic structure of the power-Maxwell black
holes}
\author{Ahmad Sheykhi}
\email{ashykhi@shirazu.ac.ir}
\affiliation{Physics Department and Biruni Observatory, College of Sciences, Shiraz
University, Shiraz 71454, Iran}
\affiliation{Research Institute for Astronomy and Astrophysics of Maragha (RIAAM), P.O.
Box 55134-441, Maragha, Iran}
\author{Mohammad Arab}
\email{m.arab92@basu.ac.ir} \affiliation{Physics Department and
Biruni Observatory, College of Sciences, Shiraz University, Shiraz
71454, Iran}

\affiliation{Department of Physics, Faculty of Science, Bu-Ali
Sina University, Hamedan 65178, Iran}

\author{Zeinab Dayyani}
\email{ashykhi@shirazu.ac.ir} \affiliation{Physics Department and
Biruni Observatory, College of Sciences, Shiraz University, Shiraz
71454, Iran}
\author{Amin Dehyadegari}
\email{ashykhi@shirazu.ac.ir} \affiliation{Physics Department and
Biruni Observatory, College of Sciences, Shiraz University, Shiraz
71454, Iran}

\begin{abstract}
Employing a new approach toward thermodynamic phase space, we
investigate the phase transition, critical behavior and
microscopic structure of higher dimensional black holes in an
Anti-de Sitter (AdS) background and in the presence of
Power-Maxwell field. In contrast to the usual extended $P-V$ phase
space where the cosmological constant (pressure) is treated as a
thermodynamic variable, we fix the cosmological constant and treat
the charge of the black hole (or more precisely $Q^s$) as a
thermodynamic variable. Based on this new standpoint, we develop
the resemblance between higher dimensional nonlinear black hole
and Van der Waals
liquid-gas system. We write down the equation of state as $%
Q^s=Q^s(T,\psi)$, where $\psi$ is the conjugate of $Q^s$, and construct a
Smarr relation based on this new phase space as $M=M(S,P,Q^s)$, while $%
s=2p/(2p-1)$ and $p$ is the power parameter of the Power-Maxwell
Lagrangian. We obtain the Gibbs free energy of the system and find
a swallowtail behaviour in Gibbs diagrams, which is a
characteristic of first-order phase transition and express the
analogy between our system and van der Waals fluid-gas system.
Moreover, we calculate the critical exponents and show that they
are independent of the model parameters and are the same as those
of Van der Waals system which is predicted by the mean field
theory. Finally , we successfully explain the microscopic behavior
of the black hole by using thermodynamic geometry. We observe a
gap in the scalar curvature $R$ occurs between small and large
black hole. The maximum amount of the gap increases as the number
of dimensions increases. We finally find that character of the
interaction among the internal constituents of the black hole
thermodynamic system is intrinsically a strong repulsive
interaction.
\end{abstract}

\maketitle

%%%%%%%%%%%%%%%%%%%%%%%%%%%%%%%%%%%%%%%%%%%%%%%%%%%%%%%%%%%%%%%
\section{introduction}
The study of black holes thermodynamics is one of the most
important subject in gravitational physics, which was anticipated
by Bekenstein in 1973 \cite{Bekenstein:1973ur}. In complete
analogy with known non-gravitational thermodynamic systems, black
hole spacetime obeys a version of first law of
thermodynamics.\cite{Bekenstein:1973ur, Hawking:1974sw}. It can be
specified by an entropy $S$ proportional to the horizon area and
temperature $T$ proportional to the surface gravity.

Furthermore, after the advent of AdS/CFT correspondence, phase
transition has gained more attention as a thermodynamical property
of black holes in asymptotically AdS spaces.

A seminal investigation in this relevance was reported in Hawking and Page's
paper \cite{Hawking:1982dh}, where it was demonstrated that in the phase
space of AdS-Schwarzchild black hole, phase transition certainly exists.
Through the AdS/CFT (gage/gravity) duality the Hawking-Page phase transition
can correspond to the confinement/deconfinement phase transition in the dual
quark gluon plasma \cite{Witten:1998zw}. In view of this duality, the
thermodynamic phase space of charged AdS black holes exhibits first order
SBH/LBH (small black hole/large black hole) phase transition suggestive of a
Van der Waals liquid/gas phase transition (see e.g \cite%
{Chamblin:1999hg,Chamblin:1999tk, Kubiznak:2012wp,
Dehghani:2014caa}).

In most treatments of phase transition in black hole
thermodynamics, in an extended phase space, negative cosmological
constant $\Lambda$ is treated as a thermodynamic variable. In
fact, $\Lambda$ is physically thought of as a pressure and its
conjugate variable is considered as a thermodynamic variable
proportional to a volume $V$ \cite{Dolan:2010ha,
Dolan:2011xt,Dolan:2012jh}. In the past few years the various
class of black hole phase transition such as Multiple reentrant
phase transition \cite{Frassino:2014pha}, superfluid-like phase
transition \cite{Hennigar:2016xwd}, zero order phase transition
\cite{Dehyadegari:2017flm} and so on have been studied in an
extended phase space (see e.g.\cite{ Cvetic:2010jb, Urano:2009xn,
Zou:2014mha, Poshteh:2013pba, Dehghani:2016wmw, Dayyani:2016gaa,
Majhi:2016txt, Dehyadegari:2017hvd}).

The authors of \cite{Kastor:2009wy}, by using of the Smarr formula,
concluded that the mass $M$ of AdS black hole should be identified as
enthalpy $H$ rather than internal energy of the spacetime . Therefor,
regarding the cosmological constant as a pressure $P$, for a non rotating
charged black hole, the first law of thermodynamics should read,
\begin{equation*}
\, \mathrm{d} M \equiv \, \mathrm{d} H= T \, \mathrm{d} S +\Phi \, \mathrm{d}
Q+ V \, \mathrm{d} P,
\end{equation*}
where $Q$ and $\Phi$ are charge and electrical potential respectively.

Although, there are some reasons to suppose the cosmological
constant as a variable, but it is more reasonable to hold it as a
constant parameter. In fact, from the physical standpoint it is
difficult to consider the cosmological constant as a thermodynamic
variable which can take an arbitrary value. Also, in general
relativity the cosmological constant is understood as a constant
related to the zero point energy of the background spacetime. By
these motivations, the authors of \cite{Dehyadegari:2016nkd}
proposed the cosmological constant as a fixed parameter and
consider the charge of the black hole as a thermodynamic variable.
They indicated a phase transition similar to Van der Waals
liquid-gas in the black hole system in this manner.
In this alternative view, the SBH/LBH phase transition of black hole in $%
Q^2-\psi$ plane exactly correspond to the Van der Walls fluid
phase transition, whereas $\psi$ is thermodynamic variable
conjugate to the $Q^2$. Recently, the universality class and
critical properties of any AdS black hole in this alternative
approach toward the phase space have been addressed in
\cite{Dehyadegari:2018pkb}.

Besides, the study of the nonlinear electrodynamic field on the
thermodynamic and phase transition of the black holes has been
attracted a lot of attentions in the literature in the past decade \cite%
{Fernando:2006gh, Myung:2008kd, Miskovic:2008ck, Myung:2008eb,
Banerjee:2011cz, Gunasekaran:2012dq,
Hendi:2014kha,Sheykhi:2014ipa, Hendi:2015xya, Dayyani:2017fuz}.
The first one of the nonlinear electromagnetic field models is
Born-Infeld electrodynamics introduced in 1930's. Nonlinear
behavior in strong electromagnetic field such as the field in
region near a point-like charge, is suggested by Dirac in $1964$
\cite{dirac1964lectures}. Moreover, nonlinear electromagnetics can
be due to the nonlinear effects of quantum electrodynamics \cite%
{Heisenberg:1935qt,Yajima:2000kw, Delphenich:2003yw}.

Power-Maxwell invariant field as an important class of the
nonlinear
electrodynamics, was introduced in \cite%
{Hassaine:2007py,Hassaine:2008pw,Hendi:2012um,Hendi:2016usw,Zangeneh:2015wia}
. It is worth mentioning that the Lagrangian of the power Maxwell
invariant fields are invariant under the conformal transformation
$g_{\mu \nu}\longrightarrow \Omega^2 g_{\mu \nu} $, where $g_{\mu
\nu}$ is the metric tensor. In the special case linear
electromagnetic can be generated
by reducing of the power Maxwell invariant field. the authors of \cite%
{Hendi:2012um} have investigated the effect of power Maxwell field on the $%
P-V$ criticality of black holes and phase transitions in the extended phase
space.

In the present work, we would like to study the critical behavior
of the AdS black hole in the presence of power-Maxwell field via
an alternative viewpoint. It means we keep the cosmological
constant as a fixed parameter
and instead treat the charge of the black hole (or more precisely $%
\boldsymbol{Q}_{p}$) as a thermodynamic variable which can vary.
The advantages of this approach is that we do not need to extend
the thermodynamical phase space to see the critical behavior of
the system. On the other side, absorbing or emitting charged
particles may cause the change in the charge of the black hole in
reasonable manner which is more logical than varying the
cosmological constant. Phase structure and critical behavior of
AdS balck holes with linear \cite{Dehyadegari:2017flm}, and
nonlinear \cite{Dehyadegari:2017hvd} electrodynamics, Lifshitz dilaton black holes \cite%
{Dayyani:2018mmm}, where the charge of the system can vary and the
cosmological constant (pressure) is fixed have been investigated.
Recently, it was argued that this method also indeed works for
investigating the phase transition of Gauss-Bonnet black holes
\cite{GB}, which further supports the viability of this
alternative approach.

The outline of this paper is as follows. In the next section we
investigate thermodynamics of the higher dimensional AdS black
hole with considering of the effects of the nonlinear power-
Maxwell field and obtain Smarr relation by the well-known usual
thermodynamic quantities.
Moreover we introduce a new phase space and thermodynamic variables $\psi ,%
\boldsymbol{Q}_{p}$ and find the appropriate Smarr relation in
terms of this new variables. In Sec.\ref{CriticalPoints}, we write
down the equation of state according to the offered variables and
calculate the critical points and critical exponent. Also, the
analogy of the system with Van der Waals liquid gas system is
showed in this section. The Gibbs free energy diagrams are
investigated in section \ref{GibbsFreeEnergy}. The swallowtail
behavior of the Gibbs free energy represents a first-order phase
transition in the system. Finally, in section
\ref{MicroscopicStractur}, we explore microscopic structure of
thermodynamic system and calculate the Ruppeiner scalar curvature
in $4,5$ and $6$ dimensional spacetime.

%%%%%%%%%%%%%%%%%%%%%%%%%%%%%%%%%%%%%%%%%%%%%%%%%%%%%%%%%%%%%%%%%%%%%%%%%%%%%%%%%%%%%%%%%%%%%
\section{Thermodynamic of higher dimensional AdS black hole with a power
Maxwell field}

The action of Einstein gravity in $(n+1)$-dimensional spacetime coupled to a
power Maxwell field can be written as \cite{Hassaine:2008pw}
\begin{equation}
I=-\frac{1}{16\pi }\int {\,\mathrm{d}^{n+1}x\sqrt{-g}}\left[ {R}-2\Lambda
+(-F_{\mu \nu }F^{\mu \nu })^{p}\right] ,
\end{equation}%
where $R$ is the Ricci scalar, $p$ is a constant determining the
nonlinearity of the electromagnetic field, $F_{\mu \nu }=\partial {_{[\mu }}%
A_{\nu ]}$ is the electromagnetic field tensor and $A_{\mu }$ is the
electromagnetic potential. Here, $\Lambda $ is the cosmological constant
related to the AdS radius as $l^{2}=-n(n-1)/\left( 2\Lambda \right) $.

We apply a spherically symmetric and static metric of $(n+1)$-dimensional
spacetime
\begin{equation}
\,\mathrm{d}s^{2}=-f(r)\,\mathrm{d}t^{2}+\frac{\,\mathrm{d}r^{2}}{f(r)}%
+r^{2}\,\mathrm{d}\Omega _{n-1}^{2},
\end{equation}%
where $\,\mathrm{d}\Omega _{n-1}^{2}$ denotes the metric of spherical
hypersurface with volume $\omega _{n-1}$ and $f(r)$ is given by \cite%
{Hassaine:2008pw}%
\begin{equation}
f(r)=1-{\frac{m}{{r}^{n-2}}}+\frac{r^{2}}{l^{2}}+{\frac{{2}^{p}\left(
2\,p-1\right) ^{2}{q}^{2\,p}}{\left( n-1\right) \left( n-2\,p\right) }{r}^{%
\left[ {(6-2n)p-2}\right] {/}\left[ 2p-1\right] }}.  \label{metric}
\end{equation}%
The quantity $q$ is an integration constant related to the charge $Q$ of the
black hole per unite volume $\omega _{n-1}$ and one can find it by applying
the Gauss law
\begin{equation}
Q=\frac{1}{4\pi }\int {r^{n-1}(-F_{\mu \nu }F^{\mu \nu })^{p-1}F_{\mu \nu
}n^{\mu }u^{\nu }\,\mathrm{d}r},  \label{Gauss}
\end{equation}%
where $n^{\mu }=f(r)^{-1/2}dt$ and $u^{\nu }=f(r)^{1/2}dr$ are the unit
space like and time like normal to the surface, respectively, and Maxwell
invariant is $F_{\mu \nu }F^{\mu \nu }=-2F_{tr}{}^{2}$. Directly from the
generalized Maxwell equation, nonzero electromagnetic field is $%
F_{tr}=q\,\,r^{{(1-n)}/{(2p-1)}}$ \cite{Hassaine:2008pw}. Thus, Gauss law
relation Eq. \eqref{Gauss} reads
\begin{equation}
Q=\frac{2^{p-1}q^{2p-1}}{4\pi }.  \label{QQ}
\end{equation}

The parameter $m$ is known as the geometrical mass parameter which it can be
expressed in term of the largest horizon radius $r_{+}$ (the largest root of
$f(r_{+})=0$)%
\begin{equation}
m={r}_{+}^{n-2}+\frac{{r}_{+}^{n}}{l^{2}}+{\frac{{2}^{p}\left( 2\,p-1\right)
^{2}{q}^{2\,p}}{\left( n-1\right) \left( n-2\,p\right) }{r}_{+}^{\left(
n-2p\right) /\left( 1-2p\right) }}.  \label{littelmass}
\end{equation}

Using Brown-York method \cite{Brown:1992br}, the total mass of the black
hole per unit volume $\omega _{n-1}$ can be read as follows%
\begin{equation}
M={\frac{\left( n-1\right) m}{16\pi }}.  \label{Mass}
\end{equation}

The electric potential $\Phi $, measured at infinity with respect to the
horizon radius $r_{+}$, is obtained as (details is referred to \cite%
{Gonzalez:2009nn})%
\begin{equation}
\Phi ={\frac{p\left( 2\,p-1\right) {\left( \pi \,Q\right) ^{1/\left(
2p-1\right) }}}{\left( n-2\,p\right) {2}^{\left( p-3\right) /\left(
2p-1\right) }}{r}_{+}^{{\left( 2p-n\right) /}\left( 2p-1\right) }}.
\label{elecpotential}
\end{equation}

\bigskip In the case $p=1$, the metric function Eq. (\ref{metric}) and the
electric potential Eq. (\ref{elecpotential}) reduce to $n+1$-dimensional
Reissner-Nordstrom (RN)-AdS black holes \cite{Kubiznak:2012wp}.

\ According to the so-called area law of the entropy, the entropy of the
black hole is a quarter of the event horizon erea. Using this, one can
obtain the entropy of the black hole per unit volume $\omega _{n-1}$ as
\begin{equation}
S=\frac{r_{+}^{n-1}}{4}.  \label{Entropy}
\end{equation}

The Hawking temperature can be calculated as
\begin{equation}
T=\frac{f^{\prime }(r_{+})}{4\pi }=\frac{\left( n-2\right) }{4\pi r_{+}}+%
\frac{nr_{+}}{4\pi {l}^{2}}-{\frac{{2}^{p-2}\left( 2\,p-1\right) q^{2\,p}}{%
\left( n-1\right) \pi }{r}_{+}^{-\left[ 2\left( n-2\right) p+1\right] /\left[
2p-1\right] }}.  \label{Temperetur}
\end{equation}

One may obtain the generalized Smarr relation for the black hole
in the extended phase space by using the definition of total mass
$M$ \eqref{Mass}, chrge of the black hole $Q$ \eqref{QQ} and the
entropy $S$ \eqref{Entropy}. It is a matter of calculation to show
the Smarr formula is
\begin{equation}
M\left( S,Q,P\right) =\frac{\left( n-1\right) S^{\left( n-2\right) /\left(
n-1\right) }}{2^{2n/\left( n-1\right) }\pi }+\frac{PS^{n/\left( n-1\right) }%
}{n2^{2n/\left( 1-n\right) }}+\frac{\left( 2p-1\right) ^{2}Q^{2p/\left(
2p-1\right) }S^{\left( 2p-n\right) /\left[ \left( n-1\right) \left(
2p-1\right) \right] }}{\left( n-2p\right) \pi ^{1/\left( 1-2p\right) }2^{%
\left[ n\left( 3p-2\right) -7p+4\right] /\left[ \left( n-1\right) \left(
2p-1\right) \right] }},  \label{massrelation}
\end{equation}%
where $P=-\Lambda /\left( 8\pi \right) $.

One can then define the variables conjugate to $Q$, $S$ and $P$. As
mentioned before, the cosmological constant parameter is treated as pressure
$P$. So, its conjugate variable from the thermodynamic viewpoint should be
volume $V$%
\begin{equation}
V=\frac{\partial M}{\partial P}\biggl|_{S,Q}.
\end{equation}%
Likewise, the corresponding conjugate quantity of $S$ and $Q$ are
interpreted as a temperature $T$ and electric potential $\Phi $ respectively
\begin{equation}
T=\frac{\partial M}{\partial S}\biggl|_{P,Q},\,\,\,\,\,\,\,\,\,\,\,\,\,\,\,%
\,\Phi =\frac{\partial M}{\partial Q}\biggl|_{S,P}.
\end{equation}

It is easy to show that the usual Smarr mass formula can be written as
\begin{equation}
M=\frac{n-1}{n-2}\,T\,S+{\frac{\left( n-3\right) p+1}{\left( n-2\right) p}}%
Q\,\Phi -{\frac{2}{n-2}\,V\,P}.  \label{smarr}
\end{equation}%
Obviously if we set $n=3$ and $p=1$, Eq.\eqref{smarr} reduces to the
well-known Smarr relation for the $4$-dimensional Einstein-Maxwell black
holes \cite{Kubiznak:2012wp}
\begin{equation}
M=2(T\,S-\,V\,P)+\Phi \ Q.  \label{smarr0}
\end{equation}%
As expected, the obtained thermodynamic quantities satisfy the usual first
law of thermodynamics
\begin{equation}
\,\mathrm{d}M=T\,\,\mathrm{d}S+\Phi \,\,\mathrm{d}Q+V\,\,\mathrm{d}P.
\end{equation}

It is notable that the electric potential $\Phi $ must have a finite value
at infinity. This leads to the following restriction on the parameter $p$%
\begin{equation}
\frac{1}{2}<p<\frac{n}{2},  \label{cons}
\end{equation}%
which obtained it from $\left( 2p-n\right) /\left( 2p-1\right) <0$ \cite{Hassaine:2008pw}.

\subsection{Alternative phase space}
\label{NewPhaseSpace} The Van der Waals like critical behavior of
various types of AdS black holes has been studied by considering
the cosmological
constant as a thermodynamical pressure in the extended phase space\cite%
{Dolan:2010ha,Kubiznak:2012wp}. Also, $Q$-$\Phi $ plane-phase transitions of
charged AdS black hole are investigated by \cite%
{Chamblin:1999hg,Chamblin:1999tk}. Although the authors claimed that the
phase transition is similar to the Van der Waals fluid system, but the phase
transition behavior exhibits unusual Van der Waals isotherms. In this
approach, a thermodynamic response function $\left( \partial Q/\partial \Phi
\right) _{T}$ does not lead to physically relevant quantity. For more
details, we refer to Ref. \cite{Dehyadegari:2016nkd}.

The most recent work that indicates a complete similarity between the
charged AdS black hole and Van der Waals fluid system is important to
highlight \cite{Dehyadegari:2016nkd}. In this work, the cosmological
constant has been thought as a fixed parameter and instead, the square of
the charge of black hole $Q^{2}$ has been considered as a thermodynamic
independent variable, where $\Psi =1/(2r_{+})$ is the conjugate of $Q^{2}$
\cite{Dehyadegari:2016nkd}. It allows the definition of new response
function $\left( \partial Q^{2}/\partial \Psi \right) _{T}$ which is clearly
characterized stable-unstable region by its sign \cite{Dehyadegari:2016nkd}.

According to this viewpoint, we would like to offer the thermodynamic
variables allowing us to complete the analogy of higher dimensional power
Maxwell-AdS black hole with Van der Waals fluid system. Hence, we consider
the mass of the black hole as a function of $\boldsymbol{Q}_{p}$ where $%
\boldsymbol{Q}_{p}\equiv Q^{2p/(2p-1)}$ instead of the standard $Q$. \textbf{%
\ }Our motivation is that the charge of the black hole appears as $Q_{p}$\
in the expressions of $M$ Eq.(\ref{massrelation}) and $T$ \ref{Temperetur}.
Another advantage of this choice is to achieve a physically meaningful
response function $\left( \partial \boldsymbol{Q}_{p}/\partial \Psi \right)
_{T}$ . Thus, with the use of \eqref{massrelation} the conjugate of $%
\boldsymbol{Q}_{p}$ is directly written as

\begin{eqnarray}
\Psi &=&\frac{\partial M}{\partial \boldsymbol{Q}_{p}}\Big|_{S,P}
\label{psir} \\
&=&\frac{(2p-1)^{2}2^{(4-3p)/(2p-1)}}{(n-2p)\pi ^{1/(1-2p)}}r_{+}^{\left(
n-2p\right) /\left( 1-2p\right) }.  \notag
\end{eqnarray}

Now we can rewrite the corresponding Smarr mass formula in this new phase
space

\begin{equation}
M=\frac{n-1}{n-2}\,T\,S+\,{\frac{2\left( \left( n-3\right) p+1\right) }{%
\left( 2\,p-1\right) \left( n-2\right) }}\boldsymbol{Q}_{p}\,\Psi -{\frac{2}{%
n-2}\,V\,P}.  \label{smarr2}
\end{equation}%
A simple calculation shows that for $p=1$ and $n=3$, relation \eqref{smarr2}
is reduced to the Smarr formula obtained by the authors of \cite%
{Dehyadegari:2016nkd}

\begin{equation}
M=2\,(\,T\,S+\boldsymbol{Q}^{2}\,\Psi -V\,P).  \label{smarr3}
\end{equation}

Finally, the first law of thermodynamics in this new picture becomes
\begin{equation}
\,\mathrm{d}M=T\,\,\mathrm{d}S+\Psi \,\,\mathrm{d}\boldsymbol{Q}_{p}+V\,\,%
\mathrm{d}P.
\end{equation}

In what follows, we examine the critical behavior of power Maxwell black
hole in alternative phase space where the AdS radius (or $\Lambda $) is
fixed and the electric charge of black hole can vary.

%%%%%%%%%%%%%%%%%%%%%%%%%%%%%%%%%%%%%%%%%%%%%%%%%%%%%%%%%%%%%%%%%%%%
\section{equation of stats and critical point}\label{CriticalPoints}
\begin{figure*}[t]
\begin{center}
\begin{minipage}[b]{0.32\textwidth}\begin{center}
                \subfigure[~$n=4$, $p=1$]{
                    \label{fig1a}\includegraphics[width=\textwidth]{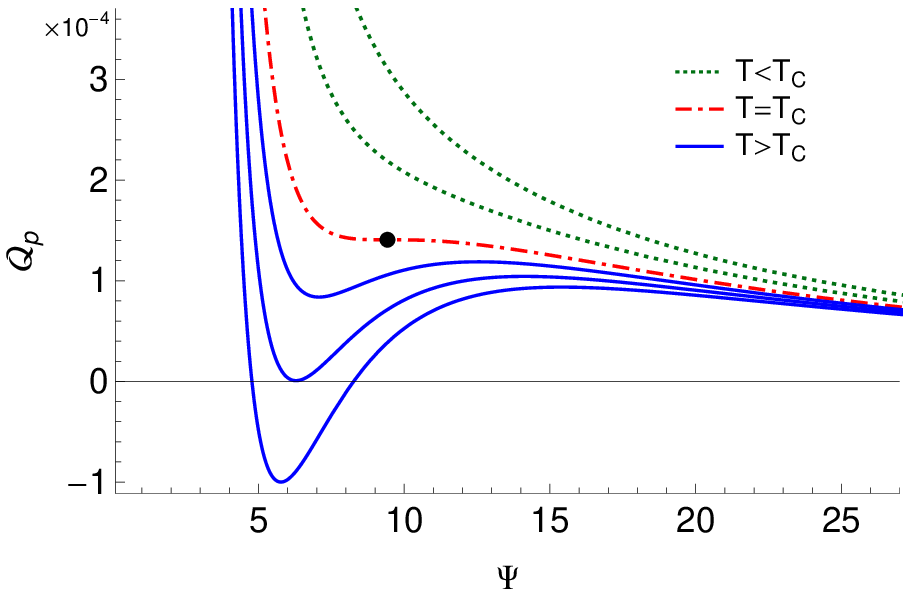}\qquad}
        \end{center}\end{minipage} \hskip+0cm
\begin{minipage}[b]{0.32\textwidth}\begin{center}
                \subfigure[~$n=4$, $p=5/4$]{
                    \label{fig1b}\includegraphics[width=\textwidth]{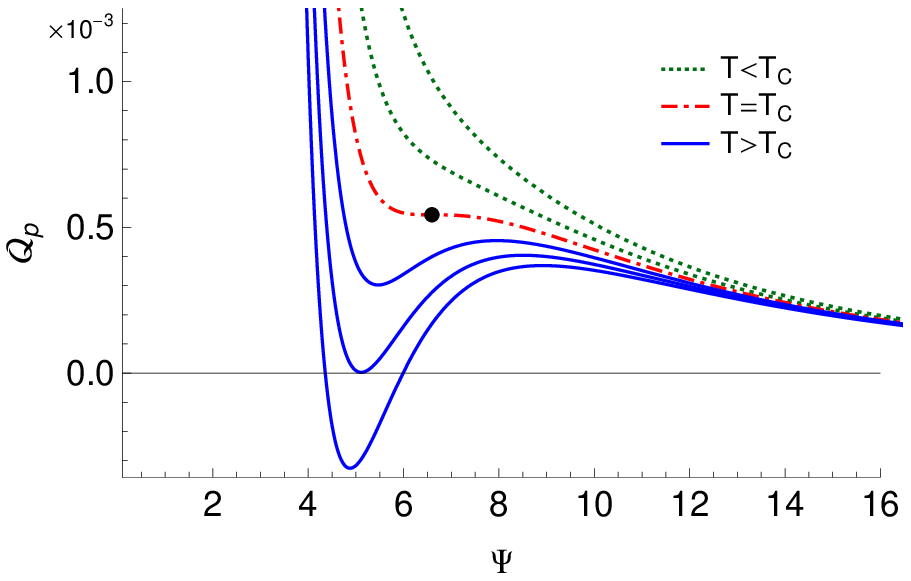}\qquad}
        \end{center}\end{minipage} \hskip0cm
\begin{minipage}[b]{0.32\textwidth}\begin{center}
                \subfigure[~$n=5$, $p=3/2$]{
                    \label{fig1c}\includegraphics[width=\textwidth]{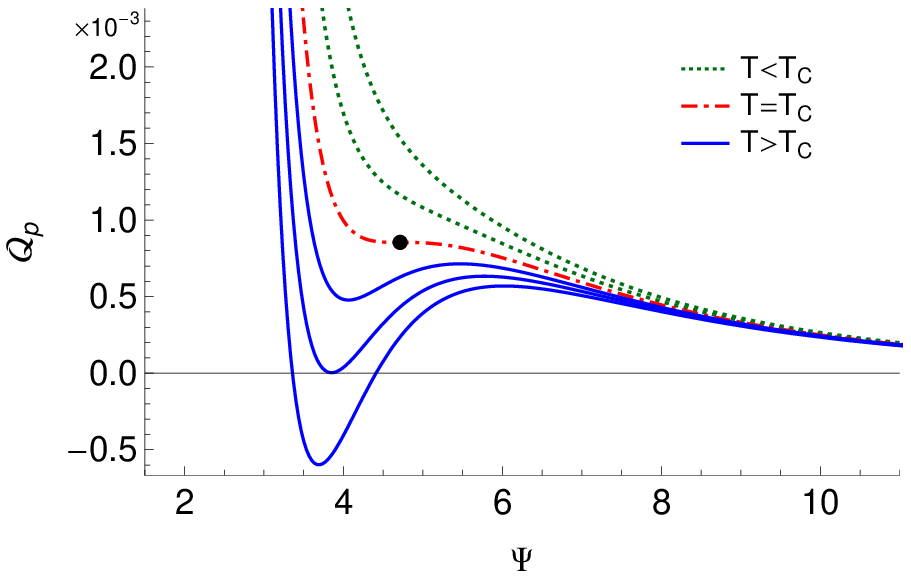}\qquad}
        \end{center}\end{minipage} \hskip0cm
\end{center}
\caption{The behavior of isothermal $\boldsymbol{Q}_{p}-\Psi $ diagrams of
power-Maxwell black holes for the case $l=1$. The critical points are
indicated by block spot.}
\label{fig1}
\end{figure*}

In order to determine the critical point, we need to have the
equation of
state $\boldsymbol{Q}_{p}\left( T,r_{+}\right) $. Using \eqref{QQ} and %
\eqref{Temperetur} one may write the equation of state as a function of the
temperature and horizon radius
\begin{equation}
\boldsymbol{Q}_{p}=\,\frac{\left( n-1\right) 2^{5p/\left( 1-2p\right) }}{%
\left( 2p-1\right) \pi ^{2p/\left( 2p-1\right) }}\left( n-2-4\pi T+\frac{%
nr_{+}^{2}}{l^{2}}\right) r_{+}^{n-3+\left( n-1\right) /\left( 2p-1\right) }.
\label{eqofstate}
\end{equation}%
The behavior of the black hole electric charge $\boldsymbol{Q}_{p}$ versus $%
\Psi $ are plotted for fixed $l=1$ and different sets of parameter values in
Fig. \ref{fig1}. In Fig. \ref{fig1}, isothermal diagrams show that, for $%
T=T_{c}$, a second-order phase transition (critical point) occurs in the
point with the following conditions (inflection point):
\begin{equation}
\frac{\partial \boldsymbol{Q}_{p}}{\partial \Psi }\Big|_{T_{c}}=0,\quad
\frac{\partial ^{2}\boldsymbol{Q}_{p}}{\partial \Psi ^{2}}\Big|_{T_{c}}=0.
\end{equation}%
Solving the above equations yields the coordinates of the critical point as
\begin{eqnarray}
\boldsymbol{Q}_{p_{c}} &=&{\frac{n\left( 2\,p-1\right) \left( n-1\right)
2^{5p/\left( 1-2p\right) }l^{n-3+\left( n-1\right) /\left( 2p-1\right) }}{%
\left( 2p\left( n-2\right) +1\right) \left( p\left( n-3\right) +1\right) \pi
^{2p/\left( 2p-1\right) }}\left[ {\frac{\left( n-2\right) \left( p\left(
n-3\right) +1\right) }{np(n-1)}}\right] }^{p(n-1)/(2p-1)},  \notag \\
\Psi _{c} &=&\frac{(2p-1)^{2}2^{\left( 3p-4\right) /(2p-1)}}{\left(
n-2p\right) \pi ^{1/\left( 2p-1\right) }}\left[ {\frac{\left( n-2\right)
\left( p\left( n-3\right) +1\right) {l}^{2}}{\left( n-1\right) np}}\right]
^{-\left[ n-2p\right] /\left[ 2\left( p-1\right) \right] },  \notag \\
T_{c} &=&{\frac{\sqrt{np\left( n-1\right) \left( n-2\right) \left( p\left(
n-3\right) +1\right) }}{\left( 2p\left( n-2\right) +1\right) \pi \,l}}.
\end{eqnarray}%
Following the new definition $\rho _{c}=\Psi _{c}\,\boldsymbol{Q}%
_{p_{c}}\,T_{c}$ in \cite{Dehyadegari:2016nkd,Dayyani:2018mmm}, the
universal number for black hole at the critical point is
\begin{eqnarray}
\rho _{c} &=&\Psi _{c}\,\boldsymbol{Q}_{p_{c}}\,T_{c}  \notag \\
&=&{\frac{n(n-1)(n-2)(2p-1)^{3}{l}^{-3+n}}{16(n-2p)(2p(n-2)+1)^{2}\pi ^{2}}}%
\left[ \frac{\left( n-2\right) (p(n-3)+1)}{np(n-1)}\right] ^{\left(
n-1\right) /2},
\end{eqnarray}%
according to constrain condition Eq.(\ref{cons}), $\rho _{c}$ is a positive
quantity. Also, it is independent of the AdS radius $l$ only when $n=3$. In
the conformally invariant case $p=\left( n+1\right) /4$, the critical
quantities of the black hole are%
\begin{eqnarray}
\boldsymbol{Q}_{c}^{c} &=&{\frac{(n-1)^{(n-1)/2}l^{n-1}}{2^{\left[ n+9\right]
/\left[ 2(n-1)\right] }\pi ^{(n+1)/(n-1)}}\left[ {\frac{\left( n-2\right) }{%
n(n+1)}}\right] }^{(n+1)/2},\quad \Psi _{c}^{c}=\frac{2^{5/(n-1)-5/2}}{l\pi
^{2/(n-1)}}\sqrt{\frac{n(n^{2}+1)}{n-2}}.  \notag \\
T_{c}^{c} &=&{\frac{1}{2\pi \,l}}\sqrt{\frac{(n+1)(n-1)(n-2)}{n}},\quad \rho
_{c}^{c}=\frac{\left( n+2/n-3\right) ^{(n+1)/2}l^{n-3}}{16\pi ^{2}\left(
n+1\right) ^{(n-1)/2}}.
\end{eqnarray}%
\begin{figure*}[t]
\begin{center}
\begin{minipage}[b]{0.32\textwidth}\begin{center}
                \subfigure[]{
                    \label{fig2a}\includegraphics[width=\textwidth]{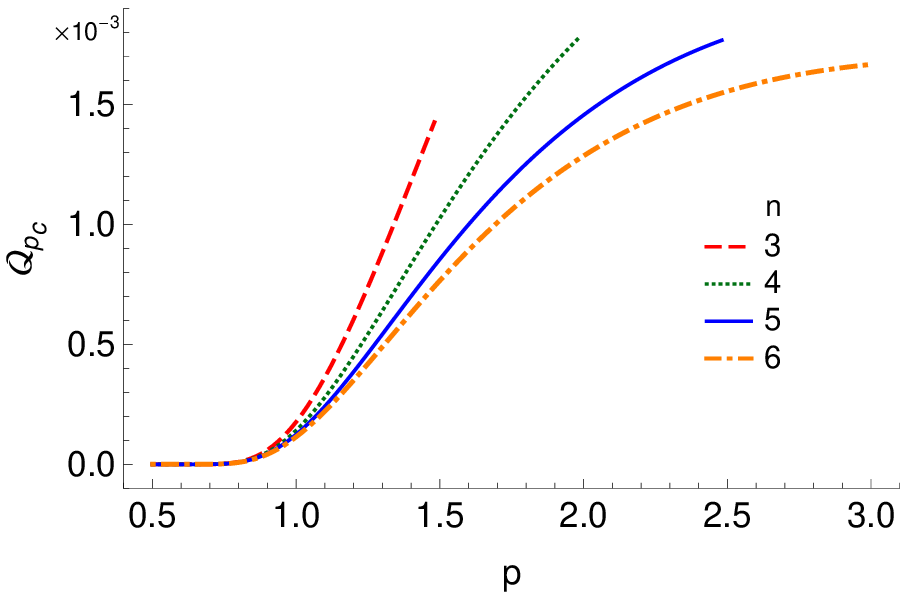}\qquad}
        \end{center}\end{minipage} \hskip+1cm
\begin{minipage}[b]{0.32\textwidth}\begin{center}
                \subfigure[]{
                    \label{fig2b}\includegraphics[width=\textwidth]{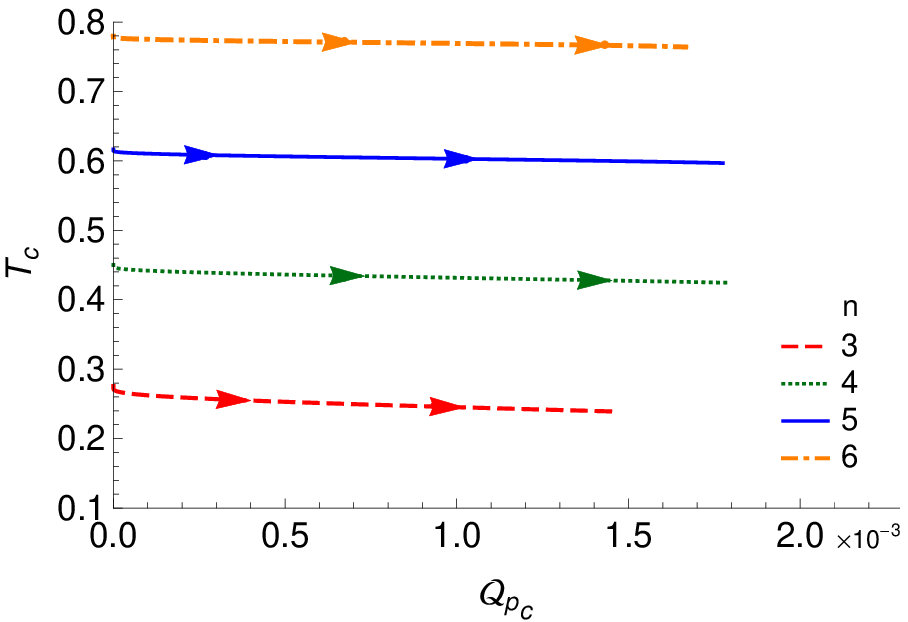}\qquad}
        \end{center}\end{minipage} \hskip0cm
\end{center}
\caption{The relation $\boldsymbol{Q}_{p_{c}}$ and $T_{c}$ and $p$ for
different values of dimensions. In (b), as $p$ increases the black hole follows
direction of arrows.}
\label{fig2}
\end{figure*}
It is remarkable to note that in $3$-dimensional space for a linear Maxwell $%
(p=1)$\textbf{,} these critical quantities reduce to those of RN-AdS black
holes \cite{Dehyadegari:2016nkd}. To see the effect of $p$ in the range of $%
1/2<p<n/2$ on the critical value of black hole, we show the behavior of $%
\boldsymbol{Q}_{p_{c}}$ and $T_{c}$ for different dimension $n$ in Fig. \ref%
{fig2}. Figure \ref{fig2a} shows, $\boldsymbol{Q}_{p_{c}}$ vanishes near $%
p=1/2$ in the different dimension. Also, increasing $p$ makes $\boldsymbol{Q}%
_{p_{c}}$higher. According to Fig. \ref{fig2b}, the critical temperature is
not almost influenced by the change of $p$.

 The existence of oscillating isotherms in Fig. \ref{fig1} are a
consequence of physically unstable feature which are remedied by the Maxwell
equal area construction \cite{Dehyadegari:2016nkd}%
\begin{equation}
\oint \Psi dQ^{2}=0.  \label{Max}
\end{equation}

\subsection{Critical exponent}

In the investigation of phase transition phenomena, it is important to study
the scaling behavior of thermodynamic system near the critical point and
find the corresponding universality class. In particular, the behavior of
physical quantities in the vicinity of the critical point can be
characterized by the critical exponents. Hence, we would like to calculate
these exponents for the new approach in this subsection.

In order to calculate the critical exponents, it is convenient to define the
reduced thermodynamic variables%
\begin{equation*}
T_{r}=\frac{T}{T_{c}},~~~~~~~~\Psi _{r}=\frac{\Psi }{\Psi _{c}},~~~~~~~%
\boldsymbol{Q}_{p_{r}}=\frac{\boldsymbol{Q}_{p}}{\boldsymbol{Q}_{p_{c}}}.
\end{equation*}%
Also, since the critical exponents should be studied near the critical
point, we rewrite the reduced variables in the form of
\begin{equation}
t=T_{r}-1,\qquad \psi =\Psi _{r}-1,\qquad \phi =\boldsymbol{Q}_{p}-1,
\end{equation}%
\textbf{\ }where $t$, $\psi $ and $\phi $ point out to the deviation from
critical point. Now, we approximate the equation of state (\ref{eqofstate})
around the critical point as
\begin{equation}
\phi =At+Bt\psi +Ct\psi ^{2}+D\psi ^{3}+O(t\psi ^{3},\psi ^{4}),
\label{sigma_s}
\end{equation}%
\textbf{\ }where $A,B$ and $C$ are constant quantities depend on $n$ and $p$%
, as follows%
\begin{eqnarray}
A &=&-{\frac{4p(n-1)\left( p(n-3)+1\right) }{\left( 2p-1\right) ^{2}}},\quad
B=\frac{4p(n-1)\left( p(n-3)+1\right) \left( 2p(n-2)+1\right) }{\left(
2p-1\right) ^{2}\left( n-2p\right) },  \notag \\
C &=&-\frac{2p(n-1)(p(n-3)+1)(2p(n-2)+1)(2p(n-3)+n+1)}{\left( 2p-1\right)
^{2}\left( n-2p\right) ^{2}},\quad D=-\frac{2p(n-1)(p(n-3)+1)(2p(n-2)+1)}{%
3(n-2p)^{3}}.
\end{eqnarray}%
Due to the fact that during phase transition the charge ($\boldsymbol{Q}_{p}$%
) remains constant, we have from Eq.(\ref{sigma_s})%
\begin{equation}
Bt\psi _{s}+Ct\psi _{s}^{2}+D\psi _{s}^{3}=Bt\psi _{l}+Ct\psi _{l}^{2}+D\psi
_{l}^{3},  \label{eosc}
\end{equation}%
where $\psi _{s}$, $\psi _{l}$ stand for the small and gas black hole,
respectively. On the other hand, by applying the Maxwell construction Eq.\ref%
{Max}, one obtains%
\begin{equation}
\int_{\psi _{s}}^{\psi _{l}}\psi (Bt+2Ct\psi +3D\psi ^{2})\,\mathrm{d}\psi
=0.  \label{maxc}
\end{equation}%
Equation (\ref{eosc}) and (\ref{maxc}) have a non-trivial solution given by

\begin{equation}
\psi _{l,s}=\frac{-Ct\pm \sqrt{3t(C^{2}t-3BD)}}{3D}.
\end{equation}%
So, the corresponding expression for the order parameter near the critical
point becomes

\begin{equation*}
|\psi _{s}-\psi _{l}|\sim t^{1/2}\Rightarrow \beta =\frac{1}{2}.
\end{equation*}%
This equations yields to the order parameter near the critical point
\begin{equation*}
|\psi _{s}-\psi _{l}|\sim t^{1/2}\Rightarrow \beta =\frac{1}{2}.
\end{equation*}%
The critical exponent $\gamma $ determines the behavior of the parameter $%
X_{T}$ as
\begin{equation*}
\chi _{_{T}}=\frac{\partial \Psi }{\partial \boldsymbol{Q}_{p}}\Big|_{T}\sim
|t|^{-\gamma },
\end{equation*}%
thus from \eqref{sigma_s},
\begin{equation}
\chi _{_{T}}\sim \frac{\Psi _{c}}{B\boldsymbol{Q}_{p_{c}}t}\Longrightarrow
\gamma =1.
\end{equation}%
The behavior of charge on the critical isotherm $t=0$ is also explained by
exponent $\delta $. Hence using \eqref{sigma_s} one can write $\phi =D\psi
^{3}$ and so $\delta =3$.

To find the specific heat behavior, one needs to find the critical exponent $%
\alpha $ such that,
\begin{equation*}
C_{\Psi }=T\frac{\partial S}{\partial T}\Big|_{\Psi }\sim |t|^{\alpha }.
\end{equation*}%
Since the entropy \eqref{Entropy} is independent of $t$, $C_{\Psi }=0$ and
we can conclude $\alpha =0$. The values we have found for the set of
critical exponents coincide with those obtained for Van der Waals fluid \cite{Kubiznak:2012wp}.

%%%%%%%%%%%%%%%%%%%%%%%%%%%%%%%%%%%%%%%%%%%%%%%%%%%%%%%%%%%%%%%%%%%%%%%%%%%%
\section{Gibbs free energy}

\label{GibbsFreeEnergy} Now, we investigate the black hole thermodynamics by
studying the thermodynamic potential. In particular, the Gibbs free energy
as a thermodynamic potential characterizes the globally stable state at
equilibrium process. Hence, to find the phase transition and classification
of its type, we explore the Gibbs free energy of power Maxwell black holes.
In fixed the AdS radius $l$ ($\Lambda $) regime, the Gibbs free energy is
calculated by Legendre transformation $G=M-TS$ \cite{Dehyadegari:2016nkd}.
Using \eqref{QQ}-\eqref{Mass} and \eqref{Entropy} the Gibbs free energy per
unit volume $\omega _{n-1}$ is obtained as%
\begin{equation}
G=G(\boldsymbol{Q}_{p},T)=\frac{r_{+}^{n-2}}{16\pi }-\frac{r_{+}^{n-2}}{%
16\pi l^{2}}+\frac{(2p-1)\left( 2p(n-2)+1\right) \pi ^{1/\left( 2p-1\right) }%
\boldsymbol{Q}_{p}}{\left( n-1\right) \left( n-2p\right) 2^{4+5p/\left(
1-2p\right) }},
\end{equation}%
where $r_{+}=r_{+}(\boldsymbol{Q}_{p},T)$.
\begin{figure*}[t]
\begin{center}
\begin{minipage}[b]{0.32\textwidth}\begin{center}
                \subfigure[~$n=4$, $p=1$]{
                    \label{fig3a}\includegraphics[width=\textwidth]{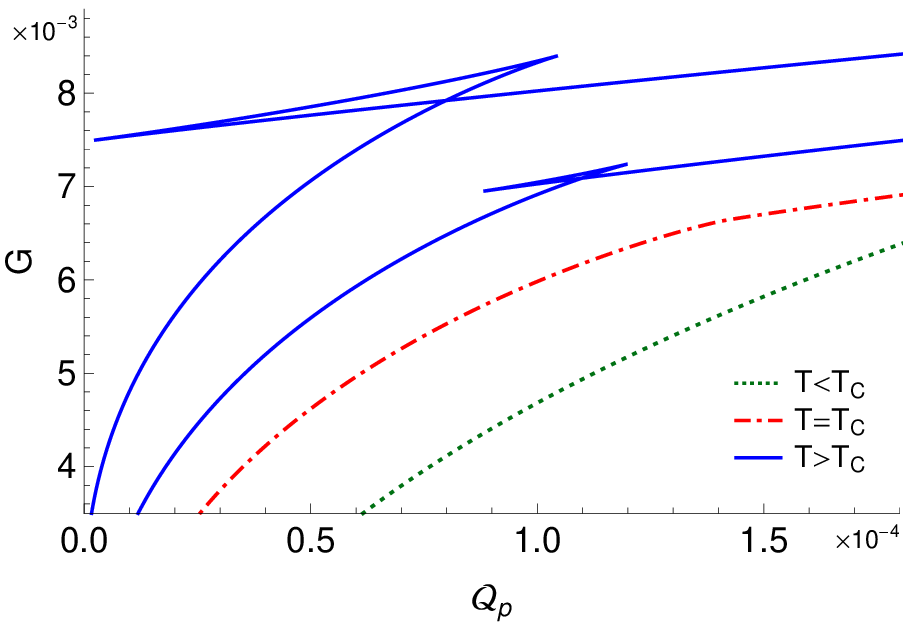}\qquad}
        \end{center}\end{minipage} \hskip+0cm
\begin{minipage}[b]{0.32\textwidth}\begin{center}
                \subfigure[~$n=4$, $p=5/4$]{
                    \label{fig3b}\includegraphics[width=\textwidth]{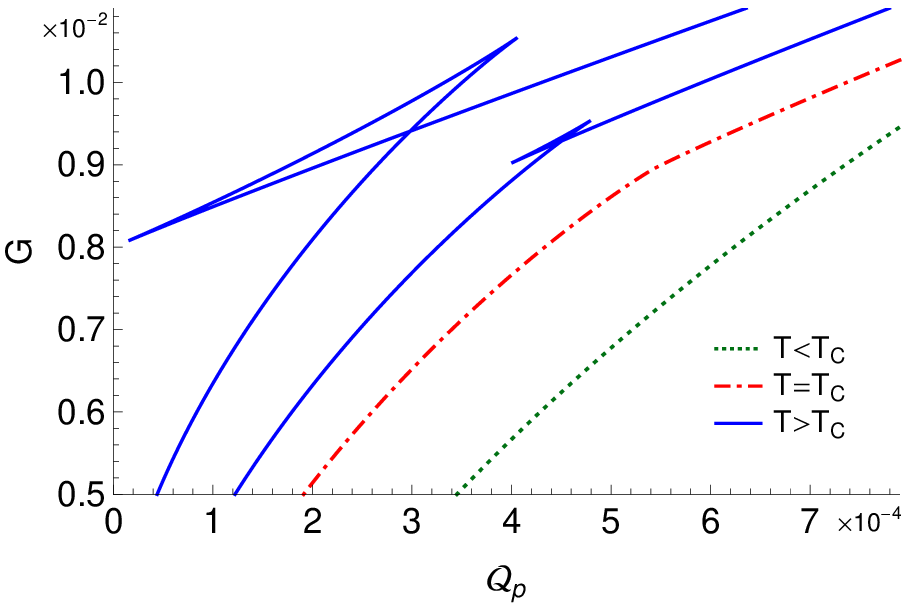}\qquad}
        \end{center}\end{minipage} \hskip0cm
\begin{minipage}[b]{0.32\textwidth}\begin{center}
                \subfigure[~$n=5$, $p=3/2$]{
                    \label{fig3c}\includegraphics[width=\textwidth]{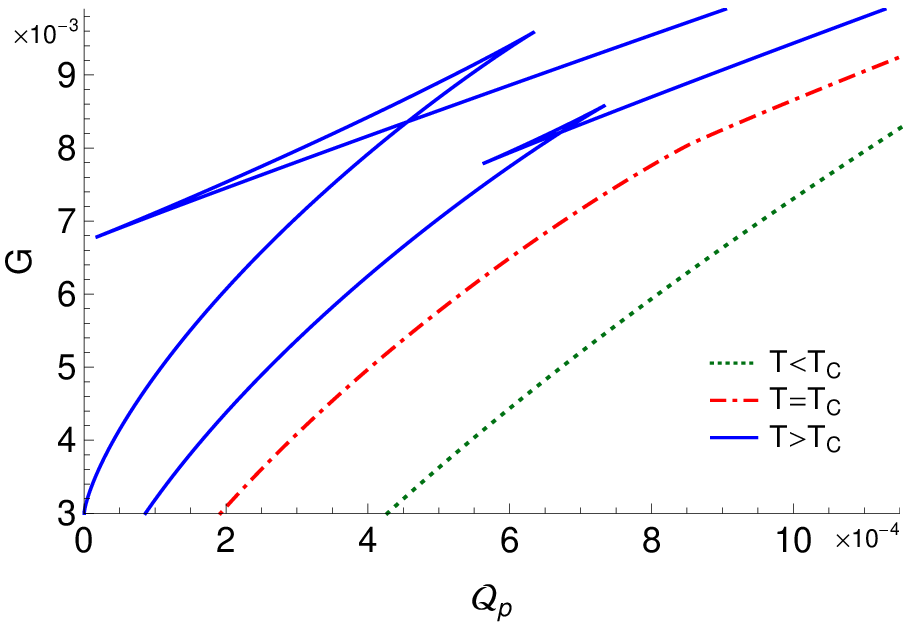}\qquad}
        \end{center}\end{minipage} \hskip0cm
\end{center}
\caption{The behavior of $G$ versus $\boldsymbol{Q}_{p}$ for power-Maxwell
black holes corresponding to Fig. \protect\ref{fig1} with $l=1$. Note curves
are shifted for clarity.}
\label{fig3}
\end{figure*}
The behavior of the Gibbs free energy $G$ is depicted in Fig. \ref{fig3}. As
it is clear from Fig. \ref{fig3}, for $T<T_{c}$, the Gibbs free energy is
single value and monotonically increasing function of $\boldsymbol{Q}_{p}$.
While for $T>T_{c}$, it becomes multivalued which means that a first-order
phase transition occurs between the small and large black holes. The
corresponding phase diagrams represented as $\boldsymbol{Q}_{p}$ versus $T$
are shown in Fig. \ref{fig4}. Here, the small and large black holes are
distinguished by transition line (blue line). As one can see from Fig. \ref%
{fig4a}, for conformal case $p=(n+1)/4$, the slope of transition line
increases with increasing dimension $n$. In Figs. \ref{fig4b} and \ref{fig4c}%
, when we increase $p$, slope of transition line increases too.

\begin{figure*}[t]
\begin{center}
\begin{minipage}[b]{0.32\textwidth}\begin{center}
                \subfigure[~$p=(n+1)/4$]{
                    \label{fig4a}\includegraphics[width=\textwidth]{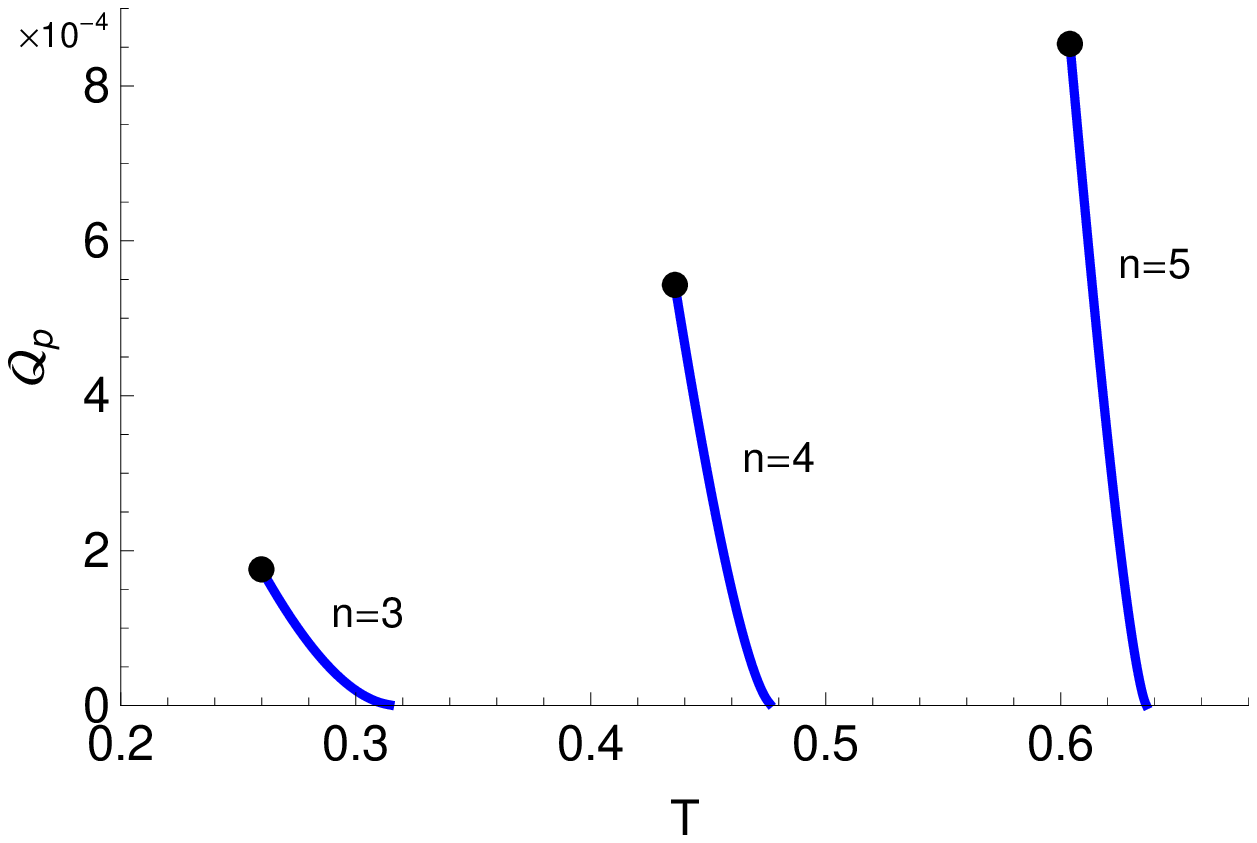}\qquad}
        \end{center}\end{minipage} \hskip+0cm
\begin{minipage}[b]{0.32\textwidth}\begin{center}
                \subfigure[~$n=3$]{
                    \label{fig4b}\includegraphics[width=\textwidth]{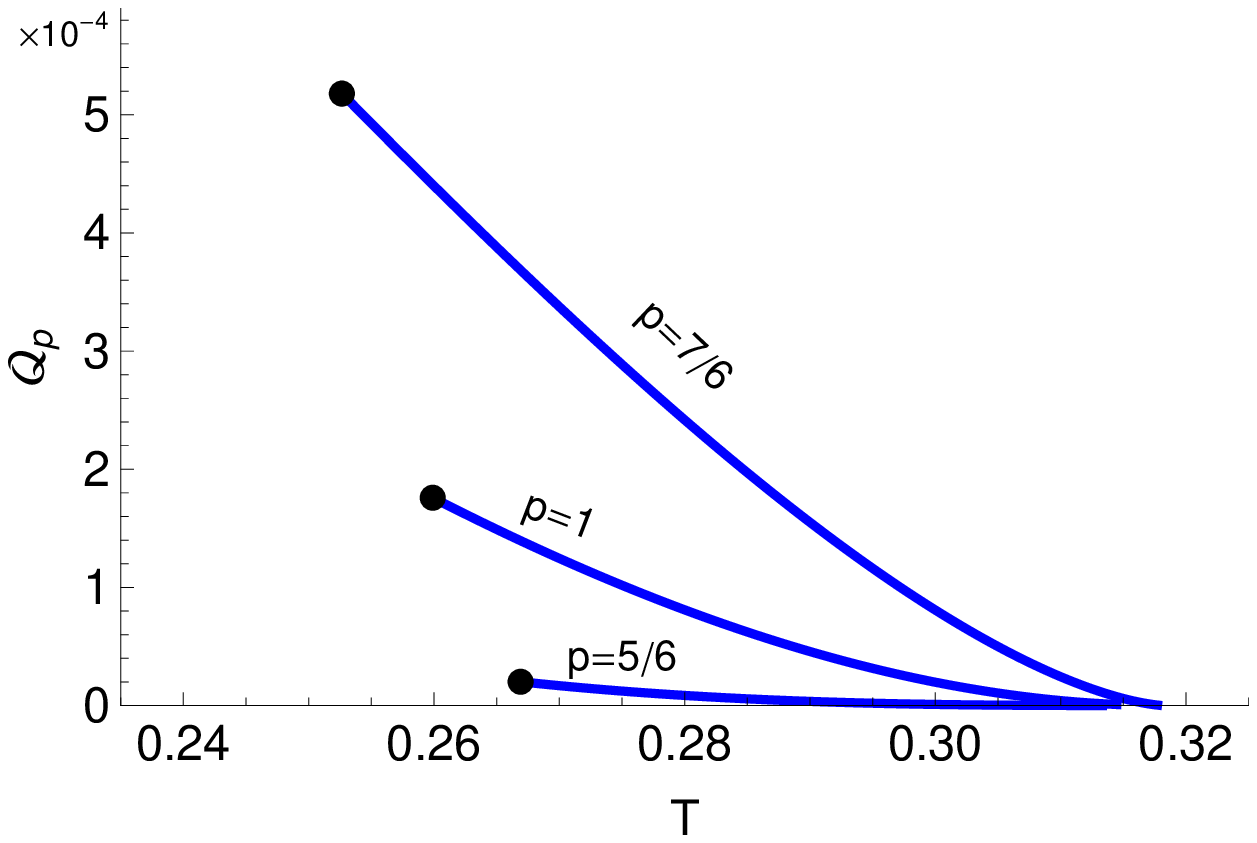}\qquad}
        \end{center}\end{minipage} \hskip0cm
\begin{minipage}[b]{0.32\textwidth}\begin{center}
                \subfigure[~$n=4$]{
                    \label{fig4c}\includegraphics[width=\textwidth]{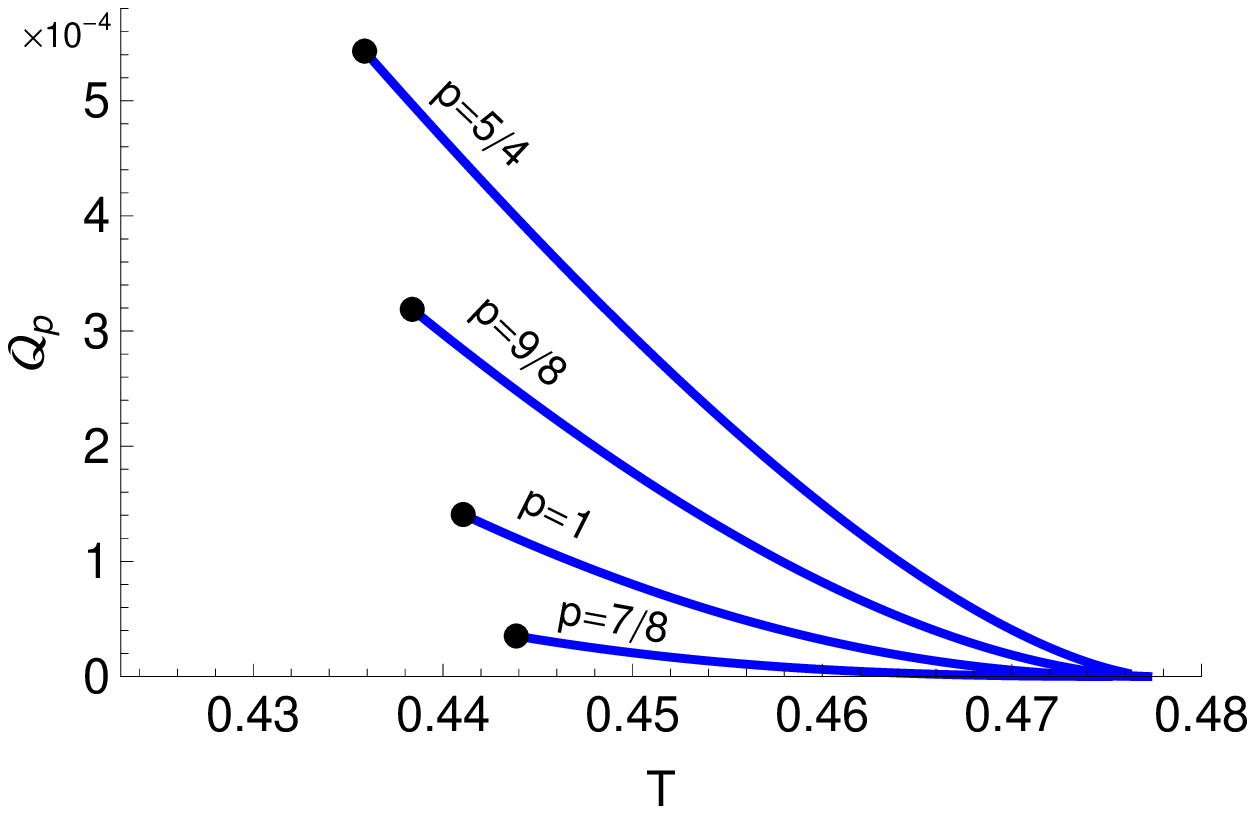}\qquad}
        \end{center}\end{minipage} \hskip0cm
\end{center}
\caption{The transition line of the phase transition between small and large
black holes for various values of $n$, $p$ and fixed $l=1$. The critical
points are indicated by block spot.}
\label{fig4}
\end{figure*}

\section{Thermodynamic geometry and microscopic structure}

\label{MicroscopicStractur}In this section, we turn to study phase
transition structure of power-Maxwell black holes in AdS space from point of
view of thermodynamic state space geometry. The Ruppeiner geometry has been
proposed as new approach to insight into underlying structure of
thermodynamic system from the thermodynamic fluctuation theory \cite{Rupp1}. Indeed,
Ricci scalar which is obtained from Ruppeiner metric, indicates the dominant
interaction between possible molecules by its sign \cite{Wei:2015iwa,comment}. In fact, the
Ruppeiner (Ricci) curvature vanishes for the ideal gas, while for attractive
(repulsive) dominant interaction is negative (positive) \cite{Ruppeiner:2008kd,may2013thermodynamic,ruppeiner2012thermodynamic}. Recently,
various studies on Ruppeiner geometry have been carried out in \cite{sahay,1610.06352,KordZangeneh:2017lgs}.

\begin{figure*}[t]
\begin{center}
\begin{minipage}[b]{0.32\textwidth}\begin{center}
                \subfigure[]{
                    \label{fig5a}\includegraphics[width=\textwidth]{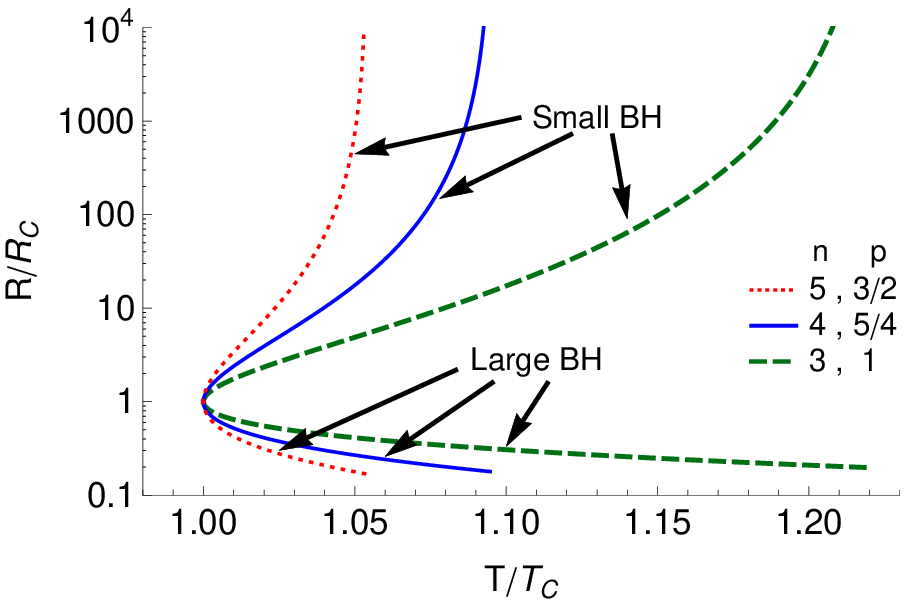}\qquad}
        \end{center}\end{minipage} \hskip+1cm
\begin{minipage}[b]{0.32\textwidth}\begin{center}
                \subfigure[]{
                    \label{fig5b}\includegraphics[width=\textwidth]{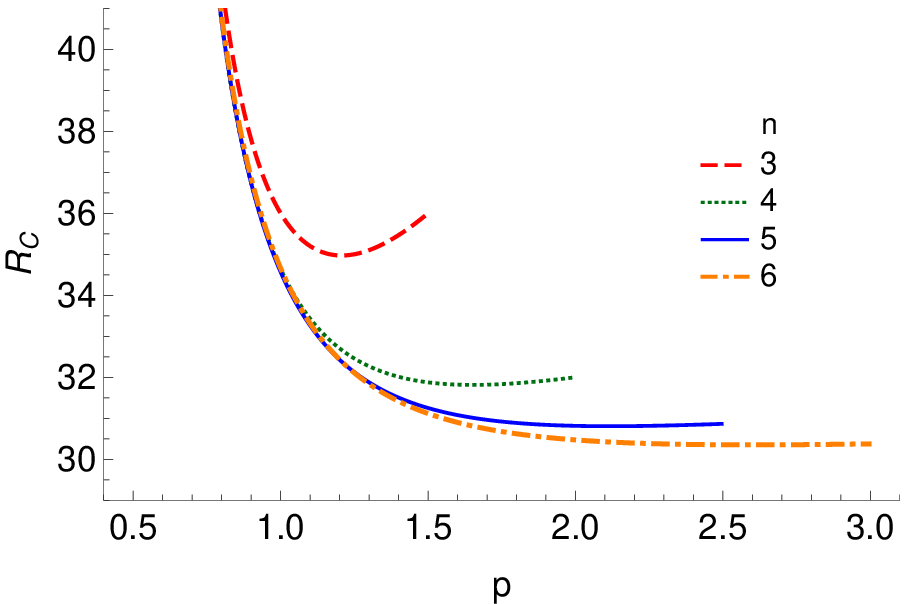}\qquad}
        \end{center}\end{minipage} \hskip0cm
\end{center}
\caption{(a) The behavior of reduced Ruppeiner scalar curvature ($R/R_{c}$)
along the transition line for large and small black holes for $p=\left(
n+1\right) /4$. (b) The dependence of the critical Ruppeiner scalar
curvature $R_{c}$ on the parameter $p$ is depicted for different values of
dimension.}
\label{fig5}
\end{figure*}

The components of the Ruppeiner metric in the energy representation are
defined as \cite{Rupp1}%
\begin{equation}
g_{\mu \nu }=\frac{1}{T}\frac{\partial ^{2}M}{\partial X^{\mu }\partial
X^{\nu }},  \label{metricRup}
\end{equation}%
where $X^{\mu }=(S,\boldsymbol{Q}_{p})$. With Eqs. (\ref{Temperetur}), (\ref%
{massrelation}) and (\ref{metricRup}) at hand, one can calculate the
Ruppeiner scalar curvature
\begin{equation}
R=\frac{8p^{\left( n+1\right) /2}\left[ \frac{(n-2)(p(n-3)+1)}{n(n-1)}\right]
^{(1-n)/2}\left[ 1+\left( \Psi /\Psi _{c}\right) ^{2(2p-1)/(n-2p)}\right]
\left( \Psi /\Psi _{c}\right) ^{\left( n+1\right) (2p-1)/(n-2p)}}{\left(
2p-1\right) \left[ 1+\frac{\left( n-1\right) }{p(n-3)+1}\left( \Psi /\Psi
_{c}\right) ^{2(2p-1)/(n-2p)}-\frac{\left( 2p-1\right) ^{2}\left(
\boldsymbol{Q}_{p}/\boldsymbol{Q}_{pc}\right) }{\left( p(n-3)+1\right)
\left( 2p(n-2)+1\right) }\left( \Psi /\Psi _{c}\right) ^{2p(n-1)/(n-2p)}%
\right] },
\end{equation}%
here $l=1$. As can be seen in Table \ref{tab}, the positive sign of $R$ is
allowed due to positive temperature, i.e. there is always repulsive
interaction. Figure \ref{fig5a} shows Ruppeiner curvature $R/R_{c}$, for
conformal case $p=(n+1)/4$, along the transition line in both the small and
large black holes. According to Fig. \ref{fig5a}, the value of Ruppeiner
curvature in both small and large black holes is the same at the critical
point. Also, there is a gap in Ruppeiner curvature between small and large
black holes that is increased in higher dimension. The dependence of the
critical Ruppeiner curvature on the allowed range of $p$ is illustrated in
Fig. \ref{fig5b}. In Fig. \ref{fig5b} for arbitrary values of dimension, the
critical value of Ruppeiner curvature diverges close to $p=1/2$.
\begin{table}[h]
\caption{The allowed ranges of $R$.}
\label{tab}%
\begin{tabular}{c|c|c|}
\cline{2-3}
& $R$ & $R $ \\ \hline
\multicolumn{1}{|c|}{$T$} & Positive & Negative \\ \hline
\multicolumn{1}{|c|}{validity} & allowed & not allowed \\ \hline
\end{tabular}%
\end{table}

%%%%%%%%%%%%%%%%%%%%%%%%%%%%%%%%%%%%%%%%%%%%%%%%%%%%%%%%%%%%%%%%%%%%%%%%%%%%%%%%%%%%%%%%%%%%%%%%%%%%%%%%%
\section{summary and conclusion}
In this paper, we have investigated the critical behavior of
higher dimensional AdS black holes with power-Maxwell nonlinear
electrodynamics via an alternative approach toward the phase
space. We have kept the cosmological constant as a fixed quantity
and treated the charge of black hole as thermodynamic variable. To
show the complete analogy between the liquid/gas phase transition
of the Van der Waals fluid and small/large black hole phase
transition, we have investigated the phase space and critical
behaviour in $\boldsymbol{Q}_{p}-\psi $ plane.

We have suggested an algorithmic method to find the
charge-independent thermodynamic
variable $\psi $ as a conjugate quantity to $f(Q)=\boldsymbol{Q}_{p}$ where $%
s={2p}/{(2p-1)}$. We have also rewritten the Smarr mass formula in
according to the new phase space and emphasized on its
correspondence with standard Smarr relation. Furthermore, we have
shown the behavior of coexistence curve of SBH and LBH in $4,5$
and $6$ dimensional spacetime. We have calculated the main
characteristic properties of the phase transition such as critical
points and critical exponent for all dimensional cases with power
Maxwell field. It was already observed that while the critical
quantities depend on the dimensions of spacetime and nonlinearity
parameter $p$, the critical exponents are independent of the
details of the system and have the same values as those of Van der
Waals system. Also, first order phase transitions are concluded
from the swallow tail behaviors of the Gibbs free energy in the
$(n+1)$-dimensional systems. It is interesting to note that with
increasing the dimensionality of the system,
the amount of transition lines gradient ($\partial \boldsymbol{Q}%
_{p}/\partial T$) is increasing.

Finally, we have studied the microscopic properties of higher
dimensional AdS black holes by considering the effects of the
conformal invariant power Maxwell field. From the viewpoint of the
thermodynamic geometry we have figured out that the interaction
between two micromolecules of black hole is a strong repulsive
interaction. Actually transition from small to large
(n+1)-dimensional black hole is due to this repulsive force.
Similar to zero temperature thermodynamic in Fermi gas, we have
introduced a temperature $T_{0}$. We have seen that at $T>T_{0}$,
when $R=0$, large black hole can be stable and its size is a
function of temperature only. Finally,
as a result, the maximum amount of the scalar curvature gap Fig. \ref{fig5a}%
{microscopic structur} has been increased by increasing the number of
dimensions.

The advantage of the approach presented in this paper is that we
do not need to extend the thermodynamical phase space to see the
Van der Wasls behaviour for black hole systems. The results
obtained here further support the argument given in
\cite{Dehyadegari:2016nkd,Dayyani:2018mmm}. Our study indicates
that the approach here is helpful to investigate the critical
behaviour of other gravity theories such as Gauss-Bonnet black
hole \cite{GB} and other different electromagnetic fields without
needing to consider the cosmological constant (pressure) as a
thermodynamic variable.

Our investigation on thermodynamical properties and microscopic
structure of the asymptotically AdS black holes with power Maxwell
field further supports the viability of this new approach and
confirms that this approach is enough powerful to explore the
phase transition of higher dimensional black holes.

%%%%%%%%%%%%%%%%%%%%%%%%%%%%%%%%%%%%%%%%%%%%%%%%%%%%%%%%%%%%%%%%%%%%%%%%%%%%%%%%%%%%%
\begin{acknowledgments}
We thank Shiraz University Research Council. The work of AS has
been supported financially by Research Institute for Astronomy and
Astrophysics of Maragha (RIAAM), Iran.
\end{acknowledgments}

%%%%%%%%%%%%%%%%%%%%
%%%%%%%%%%%%%%%%%%
%%%%%%%%%%%%%%%%%%
%%%%%%%%%%%%%%%

\end{document}